# IRCI Free Co-located MIMO Radar Based on Sufficient Cyclic Prefix OFDM Waveforms

Yun-He Cao[1], *Member*, *IEEE*, Xiang-Gen Xia[1,2], *Fellow*, *IEEE*, and Sheng-Hua Wang[1]

**Abstract**

In this paper, we propose a cyclic prefix (CP) based MIMO-OFDM range reconstruction method and its corresponding MIMO-OFDM waveform design for co-located MIMO radar systems. Our proposed MIMO-OFDM waveform design achieves the maximum signal-to-noise ratio (SNR) gain after the range reconstruction and its peak-to-average power ratio (PAPR) in the discrete time domain is also optimal, i.e., 0dB, when Zadoff-Chu sequences are used in the discrete frequency domain as the weighting coefficients for the subcarriers. We also investigate the performance when there are transmit and receive digital beamforming (DBF) pointing errors. It is shown that our proposed CP based MIMO-OFDM range reconstruction is inter-range-cell interference (IRCI) free no matter whether there are transmit and receive DBF pointing errors or not. Simulation results are presented to verify the theory and compare it with the conventional OFDM and LFM co-located MIMO radars.

This work was supported in part by the National Natural Science Foundation of China (61372136, 61172137), the Fundamental Research Funds for the Central Universities (K5051202005, K5051302089), the Air Force Office of Scientific Research (AFOSR) under Grant FA9550-12-1-0055, and by the China Scholarship Council (CSC) when Yunhe Cao was visiting the University of Delaware, Newark, DE 19716, USA.

[1] National Laboratory of Radar Signal Processing, Xidian University, Xi'an, P.R. China, 710071. (E-mail: cyh_xidian@163.com; xxia@ee.udel.edu; wshh_2011@163.com).

[2] Department of Electrical and Computer Engineering, University of Delaware, Newark, DE 19716, USA. (Email: xxia@ee.udel.edu).

# I. INTRODUCTION

Recently, there have been considerable interests in multiple-input multiple-output (MIMO) radar with multiple transmit and multiple receive antennas [1-16]. Unlike the conventional phased-array radar, MIMO radar transmits multiple orthogonal or orthogonal-like waveforms from multiple transmit antennas. MIMO radar is generally categorized into two types based on the distance between radar antennas, namely distributed MIMO radar [1]-[4] and co-located MIMO radar [5]-[16]. Distributed MIMO radar applies widely separated antennas to gain the spatial diversity, while co-located MIMO radar applies co-located transmit and receive antennas to improve spatial resolution. In this paper, we only consider co-located MIMO radar. Compared with the phased-array radar, co-located MIMO radar has been shown to offer many advantages such as increasing degrees of freedom and resolution [5],[6], improving parameter identifiability [7],[8], increasing sensitivity to detect slowly moving targets [9], enhancing flexibility for transmit beam pattern design [10],[11], and enhancing the capability of simultaneous tracking of multiple moving targets [12],[13].

In MIMO radar, waveform design for multiple transmitters is an important and challenging issue. Generally speaking, these transmitting waveforms should satisfy the following criteria:

**A$_1$**. To reduce the interference of waveforms, the multiple transmitting waveforms should be orthogonal [14] to each other or as orthogonal to each other as possible, despite their time delays [15].

**A$_2$**. In order to obtain a maximum work efficiency of the transmitter modules, a constant envelope [16] of a transmitted time domain waveform or a low peak-to-average power ratio (PAPR) is desired.

**A$_3$**. For improving the frequency efficiency and getting a maximized signal-to-noise ratio (SNR) [17] at the receiver, a constant envelope in frequency domain for the arrived signal at a target inside the total radar system bandwidth is also desired.

**A$_4$**. In order not to reduce the range resolution compared to a single transmitter radar or phased-array radar with single transmitting waveform, multiple transmitting waveforms should share the same frequency band with the same bandwidth [12],[32].

The above criterion A$_4$ is particularly important for statistical MIMO radar, where target scattering coefficients between different transmit and receive antenna pairs from the same scatter are different. This is the reason why time domain orthogonal codes/sequences (instead of frequency division waveforms) have been proposed in the MIMO radar literature but unfortunately these codes/sequences become no longer orthogonal when there are time delays among them [15]. However, it may become less important in co-located MIMO radar where all target scattering coefficients between different transmit and receive antennas from the same scatter are the same. In this case, an equivalent single transmit waveform may be formed at the receiver after applying digital beamforming (DBF) and then as long as the equivalent waveform occupies the whole signal bandwidth, the range resolution will not be reduced. This idea will be adopted in this paper for our MIMO-OFDM waveform designs.

Orthogonal frequency division multiplexing (OFDM) has been successfully used in broadband communication systems for high speed data transmissions, where the main reason is that OFDM converts an inter-symbol interference (ISI) channel to multiple ISI free subchannels, when a sufficient cyclic prefix (CP) is added to every OFDM block. In recent years, OFDM signals/waveforms have also been studied for radar applications, see for example, [18]-[32]. In most of the OFDM radar applications, OFDM waveforms do not include CP (or long enough CP) and are just treated as a different kind of radar waveforms, and at the receiver, the matched filtering is used for the range compression, see for example [18]-[26]. With this way, the key feature of converting an ISI channel to ISI free channels of the OFDM is not explored. Recently, by adding a sufficient (or maximized length) CP, a CP based OFDM range reconstruction has been proposed in [30] and [31], where there is no inter-range cell interference (IRCI) across all the range cells in a swath, i.e., IRCI free, or in other words, ideally zero sidelobes can be achieved. This idea has been extended to statistical MIMO radar in [32]. Similar idea has appeared in [28],[29] for direction of arrival estimation (not for range reconstruction). A through comparison between OFDM in communications and OFDM in radar can be found in [30].

In this paper, we consider sufficient CP based OFDM for co-located MIMO radar. We propose an IRCI free range reconstruction algorithm for co-located MIMO-OFDM radar by combing with transmit and receive DBF. We then propose a design for OFDM waveforms used for our proposed range reconstruction. Although different OFDM waveforms at different transmit antennas occupy different subbands, i.e., non-overlapped subbands, their corresponding equivalent waveform at the receiver after the transmit and receive DBF occupies the whole bandwidth and therefore the range resolution is not reduced as mentioned earlier, and furthermore, it is flat in the discrete frequency domain, which provides the maximum signal-to-noise ratio (SNR) after the range reconstruction. In addition, our designed waveforms have the optimal 0 dB peak-to-average power ratio (PAPR) in the discrete time domain, when Zadoff-Chu sequences [33]-[35] are used as the weights on the subcarriers. We then study the effects to the proposed range reconstruction when the transmit and receive DBF have pointing errors and show that the property of the IRCI free range reconstruction is still maintained. We finally present some simulation results to verify the theory and compare our method with the conventional OFDM and linear frequency modulation (LFM) waveforms, which shows that our proposed method has better range reconstruction performance.

The rest of the paper is organized as follows. Section II introduces transmit and receive signal models for co-located MIMO radar with CP based OFDM waveform. Section III proposes the IRCI free range reconstruction, the required OFDM waveform properties, and a MIMO OFDM waveform design method. Section IV studies the influences of the transmit and receive DBF pointing errors. Section V presents some simulation results. At last, Section VI concludes this paper.

## II. CO-LOCATED MIMO RADAR TRANSMIT AND RECEIVE SIGNAL MODELS

Consider a MIMO radar system consisting of $M$ co-located linear transmit antennas and $Q$ co-located linear receive antennas, the distance from the $m$th transmit antenna to the first transmit antenna and from the $q$th receive antenna to the first receive antenna are $d_t(m)$ and $d_r(q)$, respectively. A MIMO radar transmitting OFDM waveforms with CP is shown in Fig.1. It can be seen that a MIMO radar transmitting CP based OFDM waveforms includes the following steps: 1) generate $M$ different complex-valued weighting sequences of length $N$; 2) take the $N$-point inverse discrete Fourier transform (IDFT) to the $M$ weighting sequences to obtain $M$ OFDM sequences in time domain; 3) insert CP of length $L$-1 to every OFDM sequence; 4) convert $M$ digital sequences to $M$ analog OFDM waveforms; 5) transmit the $M$ cyclic prefixed OFDM waveforms at $M$ transmit antennas.

The analog OFDM waveform to transmit at the $m$th transmit antenna can be written as

$$s_m(t) = \mathrm{Re}\{\frac{1}{\sqrt{N}}\sum_{k=0}^{N-1}U_m(k)e^{j2\pi k\Delta f t}e^{j2\pi f_c t}rect(\frac{t}{T+T_{cp}})\},$$ (1)

where $\mathrm{Re}\{\}$ denotes the real part, $U_m(k)$ is the complex weight transmitted over subcarrier $k$ and antenna $m$, $N$ denotes the number of subcarriers, $\Delta f = 1/T$ represents the frequency difference between two adjacent subcarriers, the bandwidth of the signal is $B = N\Delta f$, $f_c$ is the transmitting carrier frequency, $T$ and $T_{cp}$ are the OFDM symbol and the CP guard interval lengths, respectively. The head part of $s_m(t)$ for $t \in [0, T_{cp}]$ is the same as the tail part of $s_m(t)$ for $t \in (T, T+T_{cp}]$, i.e., a CP.

The complex envelope, i.e., the baseband signal, of the $m$th transmit antenna can be written as

$$u_m(t) = \frac{1}{\sqrt{N}}\sum_{k=0}^{N-1}U_m(k)e^{j2\pi k\Delta f t}rect(\frac{t}{T+T_{cp}})$$ (2)

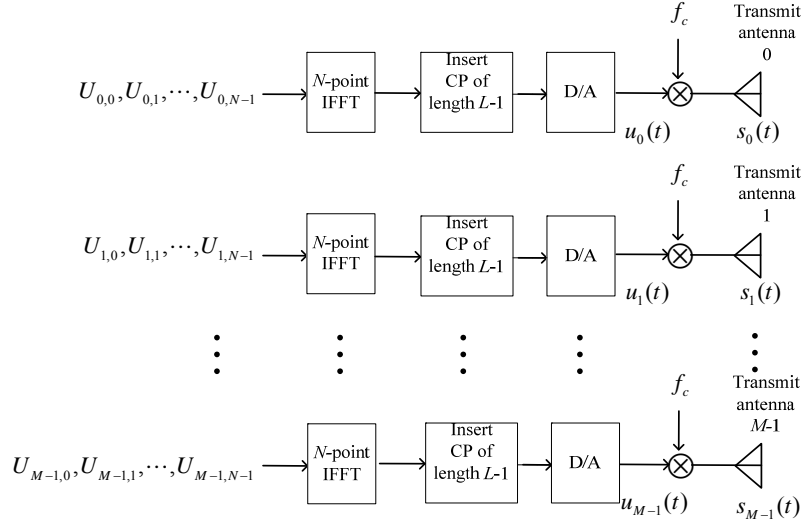

Fig.1. MIMO radar transmitters diagram.

and we have

$$s_m(t) = \mathrm{Re}\{u_m(t)e^{j2\pi f_c t}\} . \qquad (3)$$

Suppose the length of a target is $L_t$. Then, the maximal occupying range cell number of the target is

$$N_L = \left\lceil \frac{L_t}{\Delta R} \right\rceil , \qquad (4)$$

where $\lceil . \rceil$ denotes the ceiling, $\Delta R = c/(2B)$ is the range resolution, $c$ is the light propagation speed. In target tracking stage, a target has been limited to a small range area. Suppose that the tracking zone contains $L$ range cells and $L$ should satisfy $L \geq N_L$ (in practice, one may choose $L > N_L$ for tolerating possible range measurement errors). A MIMO radar transmit and receive array diagram is shown in Fig. 2. The received signal at each receive antenna is a weighted sum of the transmitted signals. These signals are reflected from a target located at position $(\varphi, \theta)$ with $\varphi$ denoting the direction of departure (DOD) and $\theta$ denoting the direction of arrival (DOA). Thus, the received signal of the $q$th receive antenna is

$$\tilde{x}_q(t) = \sum_{l=0}^{L-1}\sum_{m=0}^{M-1} g_l s_m(t - \tau_l - \gamma_m - \beta_q) + \tilde{n}_q(t)$$

$$= \mathrm{Re}\{e^{j2\pi f_c t}(\sum_{l=0}^{L-1}\sum_{m=0}^{M-1} g_l e^{-j2\pi f_c \tau_l} e^{-j2\pi f_c \gamma_m} e^{-j2\pi f_c \beta_q} u_m(t - \tau_l - \gamma_m - \beta_q) + n_q(t))\} , \qquad (5)$$

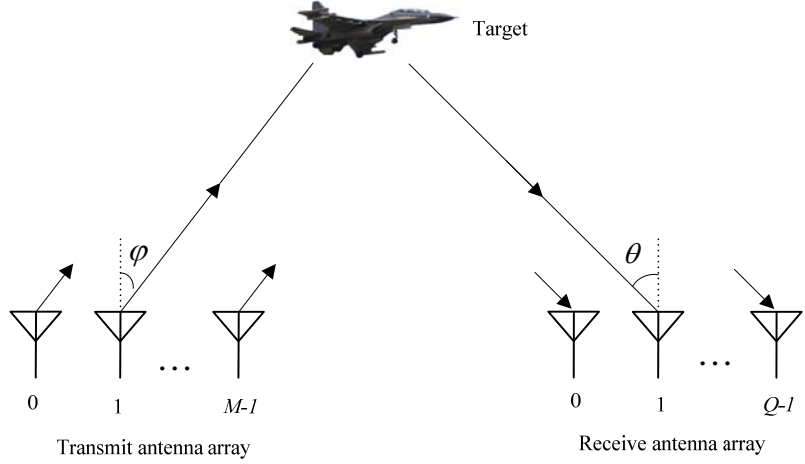

Fig. 2. MIMO radar transmit and receive diagram.

where $\tau_l = 2R_l/c$ is the delay of the $l$th range cell, $g_l$ is the radar cross section from the scattering point of the $l$th range cell. If there is no scattering point in the $l$th range cell, then $g_l = 0$. $\tilde{n}_q(t)$ is the $q$th receive antenna noise, $n_q(t)$ denotes the $q$th receive antenna noise complex envelope which is assumed to be independently and identically distributed, zero-mean complex Gaussian distribution with covariance $\sigma^2$ across both time index $t$ and spatial antenna index $q$. $\gamma_m$ and $\beta_q$ are the time delay differences in target direction from the $m$th transmit antenna to the first antenna and from the $q$th receive antenna to the first receive antenna, respectively:

$$\gamma_m = \frac{d_t(m)\sin\varphi}{c} \tag{6}$$

$$\beta_q = \frac{d_r(q)\sin\theta}{c} . \tag{7}$$

Obviously, $d_t(0) = 0$ and $d_r(0) = 0$, $\gamma_0 = 0$ and $\beta_0 = 0$. The corresponding complex envelope (baseband signal) of the received signal is

$$x_q(t) = \sum_{l=0}^{L-1}\sum_{m=0}^{M-1} g_l e^{-j2\pi f_c\tau_l} e^{-j2\pi f_c\gamma_m} e^{-j2\pi f_c\beta_q} u_m(t - \tau_l - \gamma_m - \beta_q) + n_q(t) . \tag{8}$$

Assume that the transmit and receive antenna array lengths are far less than a range cell size. Thus,

$$x_q(t) \approx \sum_{l=0}^{L-1}\sum_{m=0}^{M-1} g_l e^{-j2\pi f_c\tau_l} e^{-j2\pi f_c\gamma_m} e^{-j2\pi f_c\beta_q} u_m(t - \tau_l) + n_q(t) . \tag{9}$$

By using matrix and vector representation, the above receive signal model can be written as

$$\mathbf{x}_r(t) = [x_0(t), x_1(t), \cdots, x_{Q-1}(t)]^T$$

$$= [e^{-j2\pi f_c\beta_0}, e^{-j2\pi f_c\beta_1}, \cdots, e^{-j2\pi f_c\beta_{Q-1}}]^T y(t) + \mathbf{n}(t)$$

$$= \mathbf{A}_r(\theta) y(t) + \mathbf{n}(t) ,$$
(10)

where $[\cdot]^T$ denotes the transpose, $\mathbf{A}_r(\theta)$ is the receive steering vector and can be written as

$$\mathbf{A}_r(\theta) = [e^{-j2\pi f_c\beta_0}, e^{-j2\pi f_c\beta_1}, \cdots, e^{-j2\pi f_c\beta_{Q-1}}]^T$$

$$= [1, e^{-j\frac{2\pi}{\lambda}d_r(1)\sin\theta}, \cdots, e^{-j\frac{2\pi}{\lambda}d_r(Q-1)\sin\theta}]^T ,$$
(11)

where $\lambda = c/f_c$ is the wavelength,

$$y(t) = \sum_{l=0}^{L-1}\sum_{m=0}^{M-1} g_l e^{-j2\pi f_c\tau_l} e^{-j2\pi f_c\gamma_m} u_m(t-\tau_l)$$
(12)

is the receive baseband signal of the first receive antenna, and

$$\mathbf{n}(t) = [n_0(t), n_1(t), \cdots, n_{Q-1}(t)]^T$$
(13)

is the receive noise vector. For convenience, we suppose that the DOA angle $\theta$ of the target is accurately known at the receiver. We will consider the case when this angle has errors in Section IV later. Performing receive digital beamforming (DBF), we have

$$z(t) = \mathbf{A}_r^H(\theta)\mathbf{x}_r(t)$$

$$= Q\, y(t) + \mathbf{A}_r^H(\theta)\mathbf{n}(t)$$
(14)

$$= Q\sum_{l=0}^{L-1}\sum_{m=0}^{M-1} g_l e^{-j2\pi f_c\tau_l} e^{-j2\pi f_c\gamma_m} u_m(t-\tau_l) + v(t)$$

$$= Q\sum_{l=0}^{L-1} g_l e^{-j2\pi f_c\tau_l} \mathbf{A}_t^T(\varphi)\mathbf{u}(t-\tau_l) + v(t)$$

$$= Q\sum_{l=0}^{L-1} g_l e^{-j2\pi f_c\tau_l} b(t-\tau_l) + v(t) ,$$
(15)

where $[\cdot]^H$ denotes conjugate transpose, and

$$\mathbf{A}_t(\varphi) = [e^{-j2\pi f_c\gamma_0}, e^{-j2\pi f_c\gamma_1}, \cdots, e^{-j2\pi f_c\gamma_{M-1}}]^T$$

$$= [1, e^{-j\frac{2\pi}{\lambda}d_t(1)\sin\varphi}, \cdots, e^{-j\frac{2\pi}{\lambda}d_t(M-1)\sin\varphi}]^T \tag{16}$$

is the transmit steering vector, and

$$\mathbf{u}(t-\tau_l) = [u_0(t-\tau_l), u_1(t-\tau_l), \cdots, u_{M-1}(t-\tau_l)]^T \tag{17}$$

$$b(t-\tau_l) = \mathbf{A}_t^T(\varphi)\mathbf{u}(t-\tau_l) \tag{18}$$

and

$$v(t) = \mathbf{A}_r^H(\theta)\mathbf{n}(t) \tag{19}$$

is the noise after the receive DBF. From (15) and (18), the received signal model can be thought of as that a single transmit antenna transmits an equivalent signal $b(t)$ that is a spatial synthesis signal from all the transmit antennas. This signal $b(t)$ is called the equivalent transmit signal of the MIMO radar system.

## III. IRCI FREE RANGE RECONSTRUCTION AND CP BASED OFDM WAVEFORM DESIGN

For clarity, we first give a receive time echo diagram as Fig. 3. Considering a tracking zone of length $T_o$. The CP length $T_{cp}$ should satisfy $T_{cp} \geq T_o$, where $T_o = 2(L-1)\Delta R/c = (L-1)/B$ (note that $\Delta R = c/(2B)$ is the range resolution) is the time delay difference from the first range cell to the last range cell of the tracking zone. In order to minimize the CP length so as to reduce the unnecessary transmission energy and also for convenience, without loss of generality, we let $T_{cp} = T_o$. By adding the CP at the beginning of the waveform, one guarantees that the received signal has a full period of the transmitted waveform symbol for each range cell after removing a portion of the echo signal, i.e., the CP part in our case here, which is similar to the CP based OFDM SAR imaging in [30] and the DOA estimation in [28].

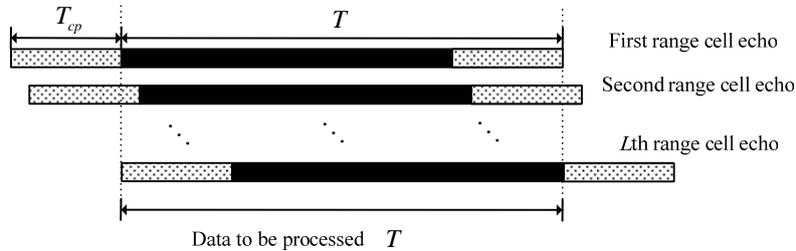

Fig.3. Target echo signals from a waveform with CP.

## A. Discrete OFDM waveform with CP

After analog to digital (A/D) sampling from the first range cell of the tracking zone with sampling frequency $f_s = B$ and the sampling interval length $T_s = 1/f_s$, $T_{cp} = (L-1)T_s$, $T = NT_s$, the discrete received signal form of (15) can be written as

$$z(n) = Q\sum_{l=0}^{L-1} h(l)b(n-l) + v(n), \quad 0 \le n < N + 2L - 2,\quad (20)$$

where, for simplicity, we assume $\tau_l = lT_s$, the complex scattering coefficient of the $l$th range cell is

$$h(l) = g_l e^{-j2\pi f_c \tau_l} \quad (21)$$

and

$$b(n) = \begin{cases} \mathbf{A}_t^T(\varphi)\mathbf{u}(n), & 0 \le n < N + L - 1 \\ 0, & else \end{cases}$$

$$= \begin{cases} \sum_{m=0}^{M-1} e^{-j2\pi f_c \gamma_m} u_m(n), & 0 \le n < N + L - 1 \\ 0, & else \end{cases} \quad (22)$$

is the discrete spatial synthesis signal of all the transmit waveforms in $\varphi$ angle where $\gamma_m = d_t(m)sin\varphi/c$.

$$\mathbf{u}(n) = [u_0(n), u_1(n), \cdots, u_{M-1}(n)]^T \quad (23)$$

is the discrete form of $\mathbf{u}(t - \tau_l)$, $u_m(n)$ is the discrete baseband signal of the $m$th transmit antenna and can be obtained by the IDFT:

$$u_m(n) = \begin{cases} \dfrac{1}{\sqrt{N}}\sum_{k=0}^{N-1} U_m(k) e^{j2\pi \frac{nk}{N}}, & 0 \le n < N + L - 1 \\ 0, & else \end{cases}. \quad (24)$$

Clearly for every $m$, $u_m(n)$ is periodic with period $N$, i.e.,

$$u_m(n) = u_m(n+N), 0 \le n < L - 1,\quad (25)$$

where $u_m(n)$ for $0 \le n < L - 1$, is the CP of the $m$th discrete time waveform (sequence) with the length $L - 1$. Since $u_m(n) = u_m(n+N)$, $0 \le n < L - 1$, for every $m$, it follows from (22) that

$$b(n) = b(n+N), \quad 0 \le n < L - 1. \quad (26)$$

Hence, $b(n)$ can be considered as a CP based OFDM signal with CP length $L$-1 for single transmitter radar as what is studied in [30].

*B. IRCI free range reconstruction*

Following the IRCI free range reconstruction algorithm in [30] for single transmit CP based OFDM radar, we remove the CP part in the discrete time received signal model in (20) by taking the $N$ samples starting from the ($L-1$)th sample point:

$$\overline{z}(n) = z(n + L - 1), \quad 0 \leq n < N ,\tag{27}$$

where the discrete time interval of length $N$ corresponds to the analog time interval $[T_{cp}, T + T_{cp}]$ with $T_{cp} = (L-1)T_s$ of length $T$ as illustrated in Fig. 3. Then,

$$\overline{z}(n) = Q\sum_{l=0}^{L-1} h(l)b(n + L - 1 - l) + v(n + L - 1), \quad 0 \leq n < N\tag{28}$$

and the $N$-point DFT of $\overline{z}(n)$ becomes

$$\overline{Z}(k) = Q\sqrt{N}H(k)B(k)e^{j2\pi\frac{(L-1)k}{N}} + V(k) ,\tag{29}$$

where

$$H(k) = \frac{1}{\sqrt{N}}\sum_{l=0}^{L-1} h(l)e^{-j2\pi\frac{lk}{N}} = \frac{1}{\sqrt{N}}\sum_{l=0}^{N-1}\overline{h}(l)e^{-j2\pi\frac{lk}{N}} ,\tag{30}$$

where $\overline{h}(l) = \begin{cases} h(l), & 0 \leq l < L \\ 0, & L \leq l < N \end{cases}$, and

$$B(k) = \frac{1}{\sqrt{N}}\sum_{n=0}^{N-1} b(n)e^{-j2\pi\frac{nk}{N}} ,\tag{31}$$

$$V(k) = \frac{1}{\sqrt{N}}\sum_{n=0}^{N-1} v(n + L - 1)e^{-j2\pi\frac{nk}{N}}\tag{32}$$

are the $N$-point DFTs of $b(n)$ and $v(n+L\text{-}1)$, respectively.

For convenience, we assume that the DOD angle $\varphi$ of the target is accurately known at the receiver. We will consider the case when this angle has error in Section IV. In this case, $B(k)$ is known accurately at the receiver. Then, from (29), $H(k)$ can be estimated as

$$\hat{H}(k) = \frac{\overline{Z}(k)}{Q\sqrt{N}B(k)e^{j2\pi\frac{(L-1)k}{N}}}$$

$$= H(k) + V'(k) ,\tag{33}$$

where

$$V'(k) = \frac{V(k)B^*(k)e^{-j2\pi\frac{(L-1)k}{N}}}{Q\sqrt{N}|B(k)|^2}. \tag{34}$$

The target scattering coefficients $h(n) = g_n e^{-j2\pi f_c \tau_n}$ of all range cells can be obtained by taking the $N$-point IDFT of $\left\{\hat{H}(k)\right\}_{k=0}^{N-1}$ as

$$\hat{h}(n) = \frac{1}{\sqrt{N}}\sum_{k=0}^{N-1}\hat{H}(k)e^{j2\pi\frac{nk}{N}}$$

$$= \bar{h}(n) + v'(n)$$

$$= \begin{cases} h(n) + v'(n), & 0 \le n < L \\ v'(n), & L \le n < N \end{cases}, \tag{35}$$

where $v'(n)$ is the $N$-point IDFT of $\left\{V'(k)\right\}_{k=0}^{N-1}$. From the above range reconstruction, one can see that all the range cell scattering coefficients are recovered without any IRCI from other range cells, i.e., they are IRCI free.

### C. SNR analysis of the IRCI free range reconstruction

As aforementioned, each receive antenna noise complex envelope (baseband) $n_q(t)$ follows normal distribution $n_q(t) \sim \mathbb{CN}(0, \sigma^2)$ and is white in both time and space. With (11) and (19), we can easily get the noise distribution after receive DBF as $v(t) \sim \mathbb{CN}(0, Q\sigma^2)$. Since in the above range reconstruction (or the target scattering coefficient estimation), the $N$-point DFT and IDFT operations are mainly used and they are unitary operations, the final noise $v'(n)$ in the target scattering coefficient estimation (35) follows the following distribution

$$v'(n) \sim \mathbb{CN}(0, \frac{\sigma^2}{QN^2}\sum_{k=0}^{N-1}\frac{1}{|B(k)|^2}). \tag{36}$$

The relationship between $B(k)$ and subcarrier weights $U_m(k)$ can be obtained by applying the $N$-point DFT operation to (22):

$$B(k) = \frac{1}{\sqrt{N}}\sum_{n=0}^{N-1}b(n)e^{-j2\pi\frac{nk}{N}}$$

$$= \frac{1}{\sqrt{N}}\sum_{m=0}^{M-1}e^{-j2\pi f_c \gamma_m}\sum_{n=0}^{N-1}u_m(n)e^{-j2\pi\frac{nk}{N}}$$

$$= \sum_{m=0}^{M-1} e^{-j2\pi f_c \gamma_m} U_m(k) \ . \tag{37}$$

Using vector representations in terms of the subcarrier index $k$, we have

$$\mathbf{B} = [B(0), B(1), \cdots, B(N-1)]^T$$

$$= \sum_{m=0}^{M-1} e^{-j2\pi f_c \gamma_m} \mathbf{U}_m \ , \tag{38}$$

where $\mathbf{U}_m = [U_m(0), U_m(1), \cdots, U_m(N-1)]^T$ represents subcarrier weight vector over the $m$th transmit antenna. Without loss of generality, the transmit mean power is normalized to 1. According to the Parseval equality, this normalization is equivalent to

$$\frac{1}{MN} \sum_{m=0}^{M-1} \mathbf{U}_m^H \mathbf{U}_m = 1 \ . \tag{39}$$

Consider that $M$ vectors $\mathbf{U}_m$, $m \in [0, M-1]$, of subcarrier weights, are orthogonal[1] to each other. We then can obtain

$$\sum_{k=0}^{N-1} \left| B(k) \right|^2 = \mathbf{B}^H \mathbf{B}$$

$$= (\sum_{m=0}^{M-1} e^{j2\pi f_c \gamma_m} \mathbf{U}_m^H)(\sum_{i=0}^{M-1} e^{-j2\pi f_c \gamma_i} \mathbf{U}_i)$$

$$= \sum_{m=0}^{M-1} \mathbf{U}_m^H \mathbf{U}_m$$

$$= MN \ . \tag{40}$$

In order to minimize the noise variance in (36) of the scattering coefficient estimation or the range reconstruction in (35), we need

$$\min \sum_{k=0}^{N-1} \frac{1}{\left| B(k) \right|^2} \ . \tag{41}$$

Clearly, the above minimum is achieved when and only when

$$\left| B(k) \right|^2 = M \ , \text{ for all } k, \ \ 0 \le k < N \ . \tag{42}$$

---

[1] The advantage of this orthogonality in the discrete frequency domain is that it is not affected by time delays in time domain, while the orthogonality in time domain is sensitive to any time delays.

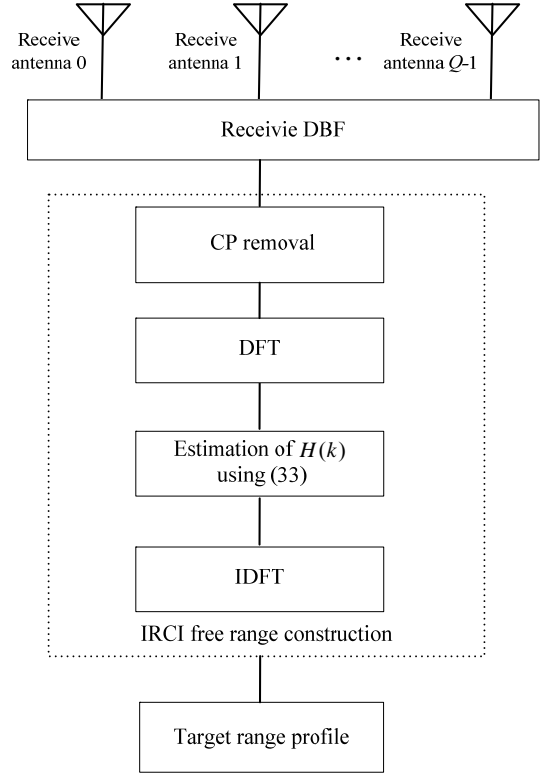

Fig. 4. MIMO radar IRCI free range reconstruction diagram.

This means that only a constant envelope $B(k)$ for all discrete frequency indices $k$ can obtain the minimum noise power or the maximum SNR of the range reconstruction in (35). Since the SNR at the $l$th range cell of the proposed algorithm is

$$SNR_{IRCI} = \frac{|h(l)|^2}{\frac{\sigma^2}{QN^2} \sum_{k=0}^{N-1} \frac{1}{|B(k)|^2}}, \tag{43}$$

when $|B(k)|^2 = M$ for every $k$, the maximal SNR of the proposed IRCI free method is achieved, which is

$$SNR_{IRCI}^{\max} = \frac{QMN |h(l)|^2}{\sigma^2}. \tag{44}$$

It can be seen that the IRCI free range processing gain is the product of the number of receive antennas, the number of transmit antennas and the gain of the matched filter (excluding CP length), i.e., the IRCI free range reconstruction method can obtain the full coherent gain of the MIMO radar system. Note that when $B(k)$ has constant module for all $k$, the range reconstruction in (33) is equivalent to the matched filtering in the frequency domain: $\overline{Z}(k)B^*(k)e^{-j2\pi\frac{(L-1)k}{N}}$. Otherwise, the range reconstruction (33) is different from the matched filtering result. The range free reconstruction is shown in Fig. 4.

### D. MIMO OFDM waveform design

From the above discussions, it is known that a constant module of the components $B(k)$ of the vector $\mathbf{B}$ is needed to maximize the range reconstruction SNR. We can see from (38) that $\mathbf{B} = \sum_{m=0}^{M-1} e^{-j2\pi f_c \gamma_m} \mathbf{U}_m$ is a weighted sum of all the subcarrier weight vectors $\mathbf{U}_m$ that are orthogonal to each other in terms of $m$ as required earlier and the weight value $e^{-j2\pi f_c \gamma_m}$ is a function of the DOD, $\varphi$, of the target. Now the question is how to design these $M$ orthogonal weight vectors $\mathbf{U}_m$ so that the components $B(k)$ of the vector $\mathbf{B}$ have constant module. Since the relationship between $\mathbf{U}_m$ and $\mathbf{B}$ depends on the DOD of a target that may change over the time, the general orthogonality between vectors $\mathbf{U}_m$ may not be good enough. In fact, this forces that the non-zero components (subcarrier weights) of $\mathbf{U}_m$ should not overlap each other for different antennas $m$, which leads to our following design for these subcarrier weight vectors $\mathbf{U}_m$. Note that, from (28) and (37), as what was mentioned earlier, $b(n)$ is an equivalent transmit signal from a single transmit antenna that arrives at the target and $B(k)$ is the $k$th discrete frequency of the equivalent transmit signal. Constant module of $B(k)$ means that the equivalent transmit signal has constant spectral power in the discrete frequency domain. This satisfies the criterion $A_3$ mentioned in Introduction and is also consistent with the single transmitter CP based OFDM radar studied in [30].

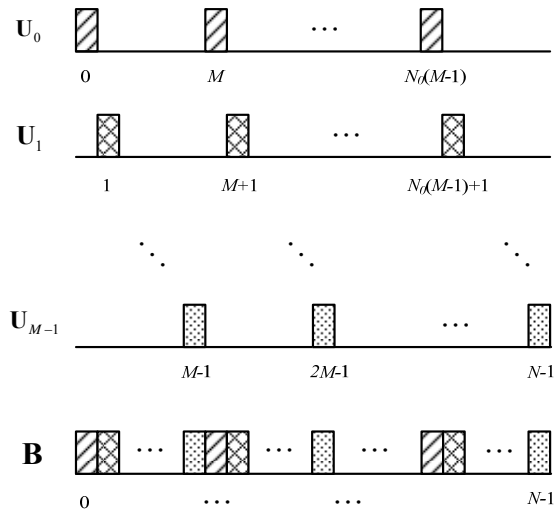

Fig.5. Interleaved structure of subcarriers for co-located MIMO OFDM radar.

To have non-overlapped weights $\mathbf{U}_m$ along the subcarriers for all the transmit antennas, there are commonly two structures: block structure and interleaved structure [27],[29]. As we shall see it later, with a block structure for $\mathbf{U}_m$ in the discrete frequency domain, it will cause the problem in designing a time domain waveform with low PAPR. In order to design OFDM waveforms with low PAPR, an interleaved structure for the weight vectors $\mathbf{U}_m$ is used, which is shown in Fig. 5. The design of $\mathbf{U}_m$ is as follows.

Without loss of generality, let us assume $N$ is a multiple of $M$, i.e., $N = N_0 M$ for some positive integer $N_0$. Let each subcarrier weight vector $\mathbf{U}_m$ has $N_0$ non-zero components with amplitude $\sqrt{M}$:

$$U_m(k) = \begin{cases} \sqrt{M}\, e^{j\phi_{m,p}}, & k = Mp + m \text{ for some integer } p \text{ with } 0 \le p < N_0 \\ 0, & else \end{cases}, \tag{45}$$

where $k \in [0, N-1]$ and the phase $\phi_{m,p}$ will be designed later when other properties are imposed to the OFDM pulses. With the above $\mathbf{U}_m$, for every $k$, $B(k)$ can be expressed as:

$$B(k) = \sqrt{M}\, e^{-j2\pi f_s \gamma_m} e^{j\phi_{m,p}}, \tag{46}$$

where $m = <k>_M$ is the remainder of $k$ modulo $M$ and $p = (k-m)/M$. Clearly, $|B(k)| = \sqrt{M}$ for all $k$, $0 \le k < N$. This means that for all the subcarrier vectors $\mathbf{U}_m$ defined in (45), the equivalent transmit signal $b(n)$ has constant module discrete spectrum $B(k)$ despite the direction of the target. The above design for the subcarrier weight vectors $\mathbf{U}_m$ does satisfy the criterion $A_4$ mentioned in Introduction. Note that although each transmit antenna only occupies $1/M$ of the signal bandwidth (one of the $M$ subbands), due to the nature of the co-located MIMO radar, their equivalent transmit signal $b(n)$ occupies the whole band and thus the range resolution is not reduced as mentioned in Introduction as well.

We next consider the time domain waveforms from the $M$ transmit antennas. Take the $N$- point IDFT to the subcarrier weights $\{U_m(k)\}_{k=0}^{N-1}$ and obtain

$$u_m(n) = \frac{1}{\sqrt{N}} \sum_{k=0}^{N-1} U_m(k) e^{j2\pi \frac{nk}{N}}$$

$$= \frac{\sqrt{M}}{\sqrt{N}} \sum_{p=0}^{N_0-1} e^{j\phi_{m,p}} e^{j2\pi \frac{n(Mp+m)}{N}}$$

$$= e^{j2\pi\frac{nm}{N}} \frac{1}{\sqrt{N_0}} \sum_{p=0}^{N_0-1} e^{j\phi_{m,p}} e^{j2\pi\frac{np}{N_0}} \tag{47}$$

Let

$$\Phi_m(n) \overset{\Delta}{=} \frac{1}{\sqrt{N_0}} \sum_{p=0}^{N_0-1} e^{j\phi_{m,p}} e^{j2\pi\frac{np}{N_0}} \tag{48}$$

Then, the module of the discrete time transmit sequence $u_m(n)$ is the same as the module of $\Phi_m(n)$, i.e., $|u_m(n)| = |\Phi_m(n)|$ for all $n$ with $0 \le n < N$. This leads us to design the phases $\phi_{m,p}$ in (45) for the subcarrier weight vectors $\mathbf{U}_m$ by using length $N_0$ Zadoff-Chu (ZC) sequences [33]-[35]. This is because if $\phi_{m,p}$ is the sequence of the phases of a ZC sequence, then, its IDFT satisfies $|\Phi_m(n)| = 1$ for all $n$, which will provide a constant module $u_m(n)$ for all $n$, i.e., the PAPR of $u_m(n)$ is 0dB, the optimum. This means that the PAPR of the discrete sequence $u_m(n)$ is low. Thus, we have the following design for the phases $\phi_{m,p}$:

$$\phi_{m,p} = -\frac{\pi}{N_0}(p + < N_0 >_2)\mu_m p, \tag{49}$$

where $\mu_m$ is a positive integer less than and relatively prime to $N_0$. The above constant module property of the discrete time signal $u_m(n)$ is due to the interleaved structure of their discrete frequency domain sequences $U_m(k)$. If blocked structures of $U_m(k)$ are used in their designs, their corresponding discrete time sequences will not have constant module. This is the reason why we have adopted the interleaved structure of $U_m(k)$ in the above design. By now, the criteria A$_2$, A$_3$, and A$_4$ are all satisfied. Although the orthogonality for the analog waveforms in time domain despite time delays from different transmit antennas may not be strictly satisfied, the discrete frequency domain orthogonality holds, which is not affected by time delays. Therefore, the orthogonality A$_1$ is also satisfied in the discrete frequency domain and ensures the IRCI free range reconstruction. By now, all the four criteria A$_1$, A$_2$, A$_3$, A$_4$ mentioned in Introduction are all satisfied for a co-located MIMO radar. Note that the above co-located MIMO-OFDM radar design also has the advantages of a co-located MIMO radar over a phased array radar and a single transmit radar, for example, it can track multiple targets simultaneously with different DODs.

## IV. INFLUENCES OF TRANSMIT AND RECEIVE DBF POINTING ERRORS

In practical radar applications, the DOD angle $\varphi$ and the DOA angle $\theta$ of a target may not be estimated very accurately and some beam pointing errors may occur. In this section, we investigate the influences, i.e., the range reconstruction SNR degradation, of these two errors.

Suppose the estimated DOA angle of the target is $\theta_0$. The receive signal in (14) after the receive DBF with this DOA angle $\theta_0$ becomes

$$\tilde{z}(t) = \mathbf{A}_r^H(\theta_0)\mathbf{x}_r(t)$$

$$= \mathbf{A}_r^H(\theta_0)\mathbf{A}_r(\theta)y(t) + \mathbf{A}_r^H(\theta_0)\mathbf{n}(t)$$

$$= \sum_{q=0}^{Q-1} e^{j2\pi f_c(\beta_q' - \beta_q)} y(t) + \tilde{v}(t)$$

$$= \tilde{Q}y(t) + \tilde{v}(t)$$

$$= \tilde{Q}\sum_{l=0}^{L-1} g_l e^{-j2\pi f_c \tau_l} b(t - \tau_l) + \tilde{v}(t) \tag{50}$$

where $\beta_q' = d_r(q)\sin\theta_0/c$, $\tilde{Q} = \sum_{q=0}^{Q-1} e^{j2\pi f_c \Delta\beta_q}$, and

$$\Delta\beta_q = \beta_q' - \beta_q = \frac{d_r(q)(\sin\theta_0 - \sin\theta)}{c}, \tag{51}$$

$$\tilde{v}(t) = \mathbf{A}_r^H(\theta_0)\mathbf{n}(t). \tag{52}$$

In this case, its Fourier domain expression (29) becomes

$$\tilde{Z}(k) = \tilde{Q}\sqrt{N}H(k)B(k)e^{j2\pi\frac{(L-1)k}{N}} + \tilde{V}(k), \tag{53}$$

where $\tilde{V}(k)$ is the $N$-point DFT of $\tilde{v}(t)$. Suppose the estimated DOD angle of the target, i.e., the angle of transmit DBF, is $\varphi_0$. Then, at the receiver, the estimated $B(k)$ in (37) becomes

$$\tilde{B}(k) = \sum_{m=0}^{M-1} e^{-j2\pi f_c \gamma_m'} U_m(k), \tag{54}$$

where $\gamma_m' = d_t(m)\sin\varphi_0/c$. In this case, the estimate of $H(k)$ in (33) becomes

$$\tilde{H}(k) = \frac{\tilde{Z}(k)}{Q\sqrt{N}\tilde{B}(k)e^{j2\pi\frac{(L-1)k}{N}}}$$

$$= \frac{[\tilde{Q}\sqrt{N}H(k)B(k)e^{j2\pi\frac{(L-1)k}{N}} + \tilde{V}(k)]\tilde{B}^*(k)e^{-j2\pi\frac{(L-1)k}{N}}}{Q\sqrt{N}\left|\tilde{B}(k)\right|^2}$$

$$= \frac{\tilde{Q}}{Q} \frac{H(k)B(k)\tilde{B}^*(k)}{\left|\tilde{B}(k)\right|^2} + \frac{\tilde{B}^*(k)e^{-j2\pi\frac{(L-1)k}{N}}}{Q\sqrt{N}\left|\tilde{B}(k)\right|^2}\tilde{V}(k) \ . \tag{55}$$

For the interleaved structure of the transmit subcarriers,

$$\left|\tilde{B}(k)\right|^2 = \sum_{m=0}^{M-1}\left|U_m(k)\right|^2 = \left|B(k)\right|^2 = M \ . \tag{56}$$

Thus, we have

$$\tilde{H}(k) = \frac{\tilde{Q}}{MQ} H(k)B(k)\tilde{B}^*(k) + \overline{V}(k) \ , \tag{57}$$

where

$$\overline{V}(k) = \frac{\tilde{B}^*(k)e^{-j2\pi\frac{(L-1)k}{N}}}{MQ\sqrt{N}}\tilde{V}(k) \ . \tag{58}$$

From (57), one can see that the range reconstruction is, in fact, still the matched filtering in the frequency domain with the estimated beam pointing from the transmit antennas. Substituting (37) and (54) into (57), we obtain

$$\tilde{H}(k) = \frac{\tilde{Q}}{MQ} H(k)\sum_{m=0}^{M-1}e^{j2\pi f_c\gamma_m'}U_m^*(k)\sum_{i=0}^{M-1}e^{-j2\pi f_c\gamma_i}U_i(k) + \overline{V}(k) \ .$$

$$= \frac{\tilde{Q}}{MQ} H(k)\sum_{m=0}^{M-1}e^{j2\pi f_c\Delta\gamma_m}\left|U_m(k)\right|^2 + \overline{V}(k) \tag{59}$$

where

$$\Delta\gamma_m = \gamma_m' - \gamma_m = \frac{d_t(m)(\sin\varphi_0 - \sin\varphi)}{c} \ . \tag{60}$$

From (45), we can write $\left|U_m(k)\right|^2$ as follows

$$\left|U_m(k)\right|^2 = \begin{cases} M, & m = <k>_M \\ 0, & \text{else} \end{cases} = \sum_{i=0}^{M-1}e^{j2\pi\frac{(k-m)}{M}i} \ . \tag{61}$$

Then, we take the $N$-point IDFT to $\tilde{H}(k)$ and obtain

$$\tilde{h}(n) = \frac{\tilde{Q}}{MQ}\sum_{m=0}^{M-1}e^{j2\pi f_c\Delta\gamma_m}\left[\frac{1}{\sqrt{N}}\sum_{i=0}^{M-1}\sum_{k=0}^{N-1}H(k)e^{j2\pi\frac{(k-m)}{M}i}e^{j2\pi\frac{nk}{N}}\right] + \overline{v}(n)$$

$$= \frac{\tilde{Q}}{MQ}\sum_{m=0}^{M-1}e^{j2\pi f_c\Delta\gamma_m}\left[\sum_{i=0}^{M-1}\frac{1}{\sqrt{N}}\sum_{k=0}^{N-1}H(k)e^{j2\pi\frac{nk}{N}}e^{j2\pi\frac{iN_0k}{N}}e^{-j2\pi\frac{mi}{M}}\right] + \overline{v}(n)$$

$$= \frac{\tilde{Q}}{MQ} \sum_{m=0}^{M-1} e^{j2\pi f_c \Delta \gamma_m} \sum_{i=0}^{M-1} h(n+iN_0) e^{-j2\pi \frac{mi}{M}} + \overline{v}(n)$$

$$= \frac{\tilde{Q}}{MQ} \sum_{i=0}^{M-1} h(n+iN_0) \sum_{m=0}^{M-1} e^{j2\pi f_c \Delta \gamma_m} e^{-j2\pi \frac{mi}{M}} + \overline{v}(n)$$

$$= \sum_{i=0}^{M-1} w_i h(n+iN_0) + \overline{v}(n) , \tag{62}$$

where

$$w_i = \frac{\tilde{Q}}{MQ} \sum_{m=0}^{M-1} e^{j2\pi f_c \Delta \gamma_m} e^{-j2\pi \frac{mi}{M}} \tag{63}$$

and

$$\overline{v}(n) = \frac{1}{\sqrt{N}} \sum_{k=0}^{N-1} \overline{V}(k) e^{j2\pi \frac{nk}{N}} . \tag{64}$$

From (63), one can see that when $\Delta \gamma_m = 0$ for all $m$, i.e., there is no DOD angle estimation error, the weights $w_i = \tilde{Q}/Q$ if $i=0$ and $w_i = 0$ otherwise. In this case, $\tilde{h}(n) = \tilde{Q} h(n)/Q + \overline{v}(n)$ in (62). When the DOA angle $\theta_0$ is also accurate, i.e., $\theta$, then $\tilde{Q} = Q$ and $\tilde{h}(n) = h(n) + \overline{v}(n)$, which coincides with what we have obtained before.

From the above derivation, we can see that when there exists a transmit DBF pointing error, no matter whether there is an error in the receive DBF pointing error or not, the target range profile is a weighted sum of $M$ range cells of the true target range profile with $N_0$ cells (one period) apart as shown in Fig. 6. In order to avoid target aliasing, the occupying range cell number of the target or tracking zone length should be less than one period

$$L < N_0 \tag{65}$$

i.e., $h(n)=0$ for $n \geq N_0$. When no noise is considered, then

$$\tilde{h}(n+iN_0) = \begin{cases} w_0 h(n), & i=0 \\ w_{M-i} h(n), & 0 \leq i < M \end{cases}, \quad 0 \leq n < N_0. \tag{66}$$

We next assume that the condition (65) holds, i.e., there is no target aliasing among the range cells. The above periodic weighting relationship leads to some disadvantages:

$\mathbf{B}_1$. The target range profile is periodic with period $N_0$ in the sense that the magnitudes of the range profile in different periods may be different, i.e., $\left| \tilde{h}(n+i_1 N_0) \right| \neq \left| \tilde{h}(n+i_2 N_0) \right|, 0 \leq n < N_0$ for $i_1 \neq i_2$.

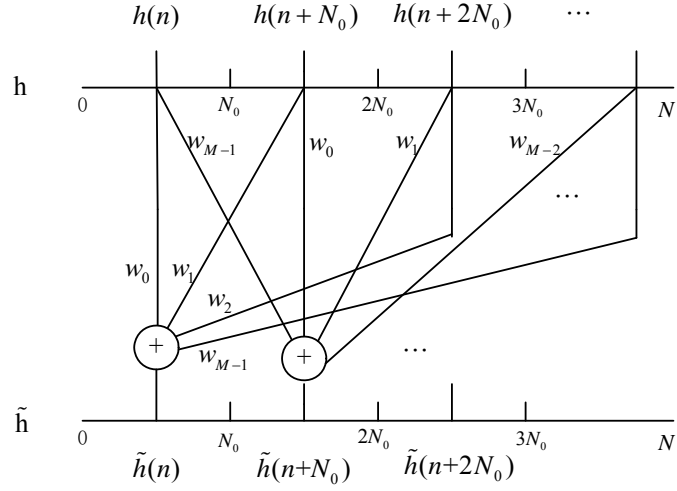

Fig. 6. Transmit beamforming output result with beam pointing error.

**B$_2$**. The SNR will decrease, because the DOD and DOA estimation errors lead to a target gain loss, i.e.,

$$\left|w_0\right| = \frac{\left|\tilde{Q}\tilde{M}\right|}{QM} < 1 \ .$$

where

$$\tilde{M} = \sum_{m=0}^{M-1} e^{j2\pi f_c \Delta \gamma_m} \ . \tag{67}$$

Since we are only interested in the target tracking zone of the first $L$ range cells, we don't need to consider the periodicity of the range profile from the transmit DBF pointing error. From (66), one can see that even when the transmit and receive DBF pointing directions have errors, our proposed range reconstruction is still IRCI free, although there may be SNR degradation as follows. The $n$th range cell coefficient $\tilde{h}(n)$ can be written as the following, when $0 \leq n < N_0$,

$$\tilde{h}(n) = w_0 h(n) + \sum_{i=1}^{M-1} w_i h(n+iN_0) + \bar{v}(n)$$

$$= w_0 h(n) + \bar{v}(n) \tag{68}$$

As aforementioned, the $N$-point DFT and IDFT operations are mainly used and they are unitary operations, the time domain noise $\bar{v}(n)$ has the following distribution

$$\bar{v}(n) \sim \mathrm{CN}(0, \frac{\sum_{k=0}^{N-1}\left|\tilde{B}^*(k)\right|^2 \sigma^2}{QM^2N^2}) \ .$$

From (56) we know that $\left|\tilde{B}^*(k)\right|^2 = \left|\tilde{B}(k)\right|^2 = M$ , then

$$\bar{v}(n) \sim \mathbb{CN}(0, \frac{\sigma^2}{QMN}).$$ (69)

The SNR at the $l$th range cell when there exist transmit and receive pointing errors is

$$SNR_{error} = \frac{|w_0|^2 |h(l)|^2}{\frac{\sigma^2}{QMN}} = \frac{\left|\tilde{Q}\right|^2 \left|\tilde{M}\right|^2 N |h(l)|^2}{QM\sigma^2},$$ (70)

The SNR loss compared to the range reconstruction when both transmit and receive DBF pointings are accurate is

$$SNR_{loss} = \frac{SNR_{IRCI}^{max}}{SNR_{error}} = \frac{Q^2 M^2}{\left|\tilde{Q}\right|^2 \left|\tilde{M}\right|^2}.$$ (71)

## V. SIMULATION RESULTS

In this section, we present some simulation results to illustrate the performance of our proposed method. We first show the performance of the proposed MIMO radar IRCI free range reconstruction with our designed OFDM waveforms. We then show the periodicity of a target profile and the SNR degradation for the IRCI free range reconstruction, when both transmit and receive DBF pointing errors occur.

### A. Performance of IRCI free range reconstruction

Suppose there are $M$=4 transmit antennas and the number of subcarriers is $N$=512. We set the tracking zone length $L$ to be 61 which is less than $N_0 = N / M = 128$ and the CP length to be $L - 1 = 60$. We also assume that a point target is located at the 40th range cell. In order to demonstrate the IRCI free property of the proposed method, we compare with the conventional OFDM waveform (no CP is added) and LFM waveform using the matched filtering. Normalized range profiles of the point spread function are shown in Fig. 7. It can be seen that the sidelobes are much lower for the CP based MIMO-OFDM signal than those of the other two signals.

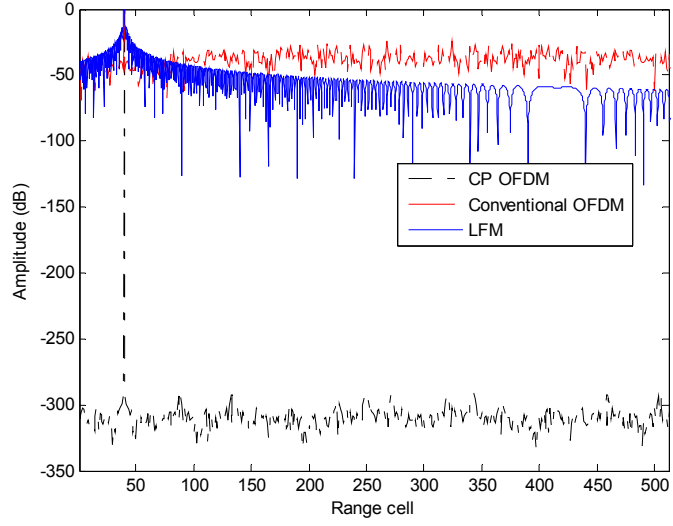

Fig. 7. Normalized range profiles of a point spread function.

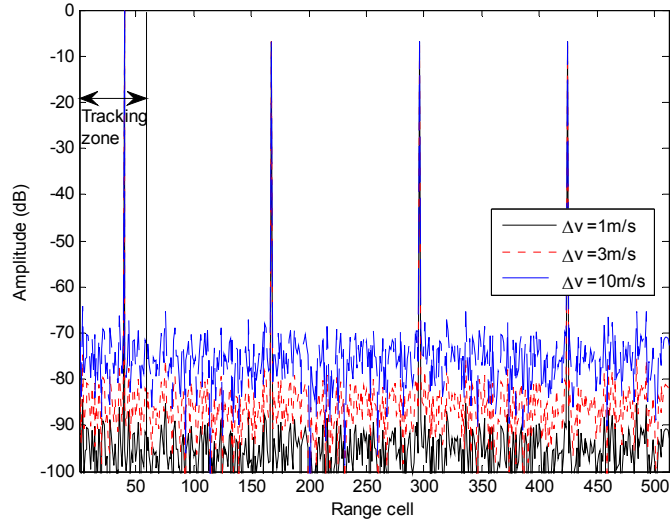

Fig. 8. Normalized range profiles of different velocity estimation error.

When the target moves and induces a Doppler in the received signal. This Doppler can be compensated at the receiver if its velocity can be estimated accurately. However, in practice the target velocity estimation may not be accurate. In this case, the range reconstruction performance may be degraded. Let us show an example for our proposed IRCI free range reconstruction method. Suppose the radar carrier frequency $f_c$ is 3GHz and the signal bandwidth is 50MHz. Normalized range profiles of the point spread function with different velocity errors are shown in Fig. 8. It can be seen that our proposed method with our newly designed waveform can still maintain low range sidelobes when there exists velocity estimation error. Note that the periodicity appeared in Fig. 8 comes from the following reason. The target motion Doppler compensation residue causes an unknown (fractional) frequency shift in the transmit DBF vector $B(k)$ in (29) that cannot be matched well by using $B(k)$ in the range reconstruction. Due to the interleaved structure of $U(k)$ in $B(k)$ in (45), (46), and (38), the residue left in (33) in the frequency domain is similar to that when there is a transmit DBF pointing error as studied in Section IV. This leads to the periodicity after the range reconstruction. Since our target tracking zone only contains the first 61 range cells that are completely contained in the first period, this periodicity does not affect the target detection.

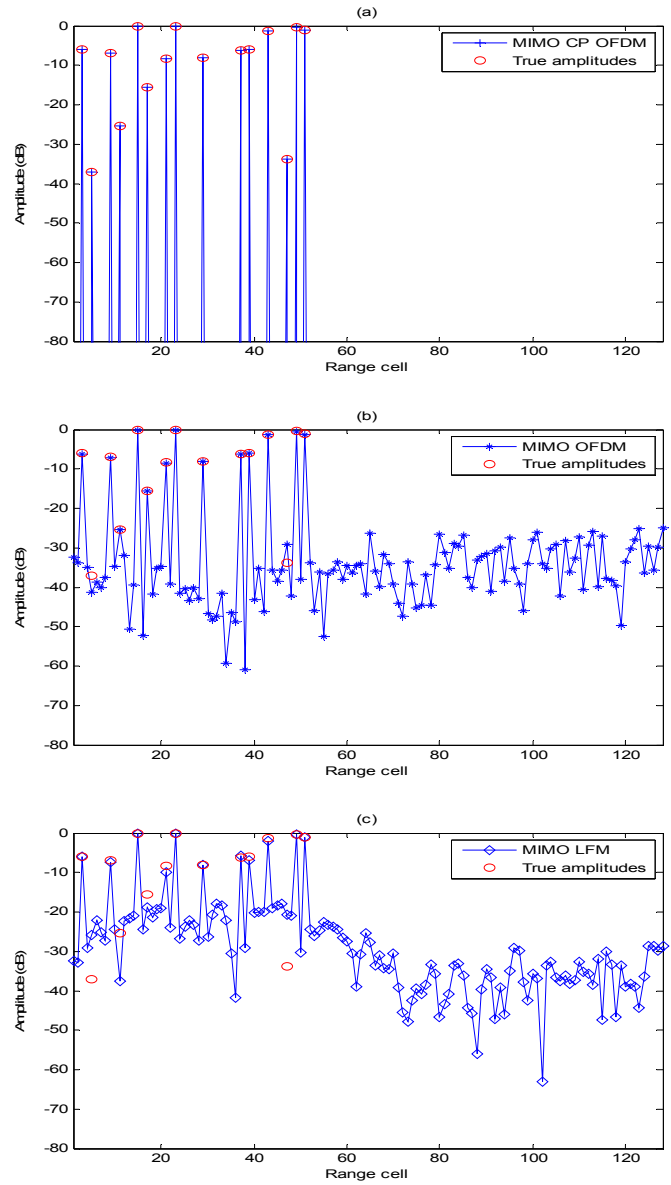

Fig. 9. Range profiles of multiple scattering points: (a) CP based OFDM waveform; (b) the conventional OFDM waveform; (c) LFM waveform.

Suppose the target spreads over several range cells with different amplitudes. Such an example is shown in Fig. 9. Since there are no IRCI between scattering points in different range cells, the range profile can be recovered perfectly. The conventional matched filtering with OFDM waveform and LFM waveform have high sidelobes and some weak scattering points are submerged by the high sidelobes of the strong scattering points.

## B. Influences of transmit and receive DBF pointing errors

Suppose co-located transmit array and receive array are uniform linear array with a half wavelength element-spacing. DOA and DOD of a target are $20^{\circ}$ and $30^{\circ}$, respectively. Assume the pointing errors of transmit and receive DBF are the same. The SNR loss in (71) against the pointing error with different antenna numbers is plotted in Fig. 10. It can be seen that a larger pointing error leads to a higher SNR loss. On the other hand, the more the antenna number is, the larger the SNR loss will be at the same pointing error.

As aforementioned, a transmit beamforming pointing error will also result in a periodic range profile. Consider $M$=4 transmit antennas and $N$=512 subcarriers, the range cell number of tracking zone is $L$=61 and the transmit beamforming pointing error is $2^{\circ}$. Assume that a target spreads over 3 range cells. It can be seen from Fig. 11 that the period is $N_0$=128 and the amplitudes are different in every period. The target range profile in target tracking zone can be reconstructed perfectly even when there exists a transmit DBF pointing error.

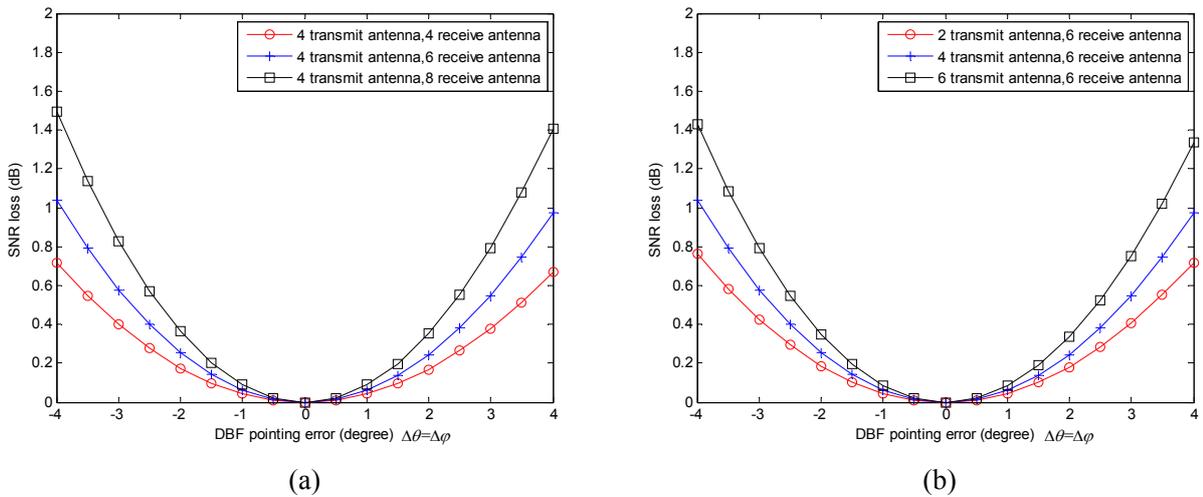

(a)                                                                (b)

Fig. 10. SNR loss against DBF pointing errors with different antenna numbers: (a) different receive antenna numbers (b) different transmit antenna numbers.

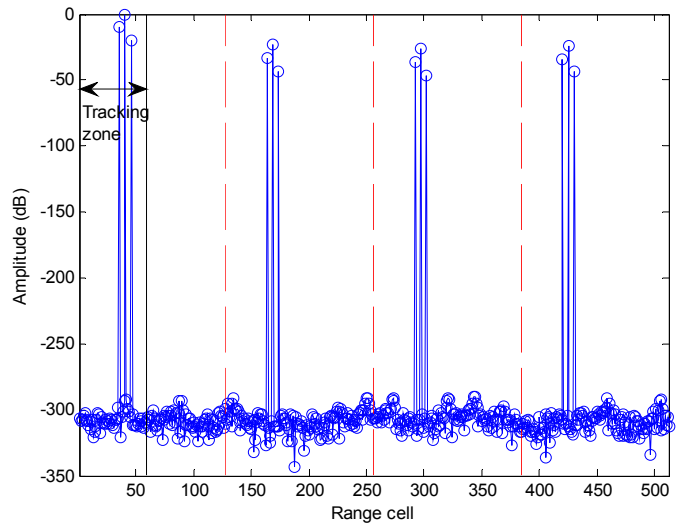

Fig. 11. Periodicity caused by transmit beamforming pointing error.

## VI. CONCLUSION

In this paper, we proposed a sufficient CP based MIMO-OFDM range reconstruction and its corresponding waveform design for co-located MIMO radar systems. Our proposed MIMO-OFDM waveform design achieves the maximum SNR gain after the range reconstruction and its PAPR in the discrete time domain is also optimal, i.e., 0dB, when Zadoff-Chu sequences are used in the discrete frequency domain as the weighting coefficients for the subcarriers. We also studied the performance when there are transmit and receive DBF pointing errors. It was shown that our proposed CP based MIMO-OFDM range reconstruction is IRCI free no matter whether there are transmit and receive DBF pointing errors or not. We finally presented some simulation results to verify the theory and compare with the conventional OFDM waveform and the LFM waveform radar.